\documentclass[twocolumn,aps,prl] {revtex4} 
\usepackage{graphicx}
\begin{document}
\pagestyle{empty} 
\title{Theory of the leak-rate of seals}
\author{  
B.N.J. Persson and C. Yang} 
\affiliation{Institut f\"ur Festk\"orperforschung, Forschungszentrum J\"ulich, D-52425 J\"ulich, Germany}

\begin{abstract}
Seals are extremely useful devices to prevent fluid leakage. However, 
the exact mechanism of roughness induced leakage is not well understood. 
We present a theory of the leak-rate of seals, which is 
based on percolation theory and a recently developed contact mechanics theory.
We study both static and
dynamics seals. We present molecular dynamics results which show that 
when two elastic solids with randomly
rough surfaces are squeezed together, as a function of 
increasing magnification or decreasing squeezing pressure, a non-contact channel will percolate 
when the (relative) projected contact area, $A/A_0$, is of order $0.4$, in accordance with 
percolation theory. We suggest a simple experiment which can be used to test the theory. 
\end{abstract}
\maketitle


{\bf 1. Introduction}

A seal is a device for closing a gap or making a joint fluid tight\cite{Flitney}.
Seals play a crucial role in many modern engineering devices, and the failure of
seals may result in catastrophic events, such as the Challenger disaster. 
In spite of its apparent 
simplicity, it is still not possible to predict theoretically the leak-rate and
(for dynamic seals) the friction forces\cite{Mofidi} for seals. 
The main problem is the influence of surface
roughness on the contact mechanics at the seal-substrate interface. 
Most surfaces of engineering interest have surface roughness
on a wide range of length scales\cite{P3}, e.g, from cm to nm, which will influence the leak rate
and friction of seals, and accounting for the whole range of surface roughness
is impossible using standard numerical methods, such as the Finite Element Method.

In this paper we will analyze the role of surface roughness on seals. We will use a recently
developed contact mechanics theory\cite{JCPpers,PerssonPRL,PSSR,P1,Bucher,YangPersson}
to calculate the leak-rate of static seals. We assume
that purely elastic deformation occurs in the solids, which is
the case for rubber seals. For metal seals, strong
plastic deformation often occurs in the contact region.

\begin{figure}
\includegraphics[width=0.45\textwidth,angle=0]{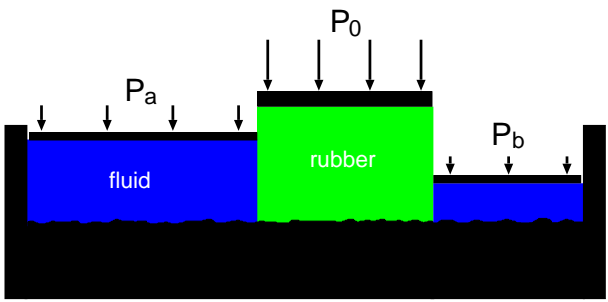}
\caption{\label{water.rubber}
Rubber seal (schematic). The liquid on the left-hand-side is under the hydrostatic
pressure $P_{\rm a}$ and the liquid to the right under the pressure $P_{\rm b}$
(usually, $P_{\rm b}$ is the atmospheric pressure).
The pressure difference $\Delta P = P_{\rm a}-P_{\rm b}$ results in liquid flow
at the interface between the rubber seal and the rough substrate surface. The
volume of liquid flow per unit time is denoted by $\dot Q$, and depends on the 
squeezing pressure $P_0$ acting on the rubber seal. 
}
\end{figure}

The theory developed below is based on studying the interface between the rubber and the
hard countersurface (usually a metal) at different magnifications $\zeta$. At low magnification the surfaces
appears flat and the contact between them appears to be complete (i.e., no leak channels can be observed). 
However, when we
increase the magnification we observe surface roughness at the interface, and, in general,
non-contact regions. As the magnification increases, we will observe more and more
(short-wavelength) roughness, and the (apparent) contact area $A(\zeta )$ 
between the solids will decrease. At 
high enough magnification, for $\zeta=\zeta_{\rm c}$, a non-contact (percolation) 
channel will appear, through which fluid will flow, from the
high pressure side (pressure $P_{\rm a}$) to the low pressure 
side (pressure $P_{\rm b}$), see Fig. \ref{water.rubber}.
We denote the most narrow passage between the two surfaces along the percolation path as the
critical constriction. When the magnification increases further, more percolation channels will be observed,
but these channels will have more narrow constrictions than those for the first channel which appears at the 
percolation threshold ($\zeta=\zeta_{\rm c}$).  

The picture described above for the leakage of seals has already been
presented by one of the present authors\cite{Creton,P3}. However, recent development in
contact mechanics now allows us to present a more accurate analysis of the leakage process.
In this paper we extend the theory of Ref. \cite{Creton}, 
and present numerical results for the size of the critical constriction
and for the leak-rate.

In Sec. 2 we describe the basic picture used to calculate the leak-rate of static seals.
In Sec. 3 we present numerical results for the size of the critical constriction and for the leak-rate.
In Sec. 4 we present Molecular Dynamics results which illustrate how the contact between
the two solids changes as the magnification $\zeta$ increases. We find that the percolation channel is formed
when $\zeta=\zeta_{\rm c}$, where $A(\zeta_{\rm c})/A_0 \approx 0.4$, 
in accordance with percolation theory\cite{Stauffer}.
In Sec. 5 we improve the theoretical picture of how to understand static seals.
In Sec. 6 we compare the theory with experimental data. In Sec. 7
we present some comments related to the non-uniform seal 
pressure distribution, the role of adhesion and rubber
viscoelasticity. In Sec. 8 we study 
dynamical (linear reciprocal motion) 
seals at low sliding velocities. In Sec. 9 we suggest a simple experiment to test
the theory. Sec. 10 contains the summary and
the conclusion.

\begin{figure}
\includegraphics[width=0.45\textwidth,angle=0]{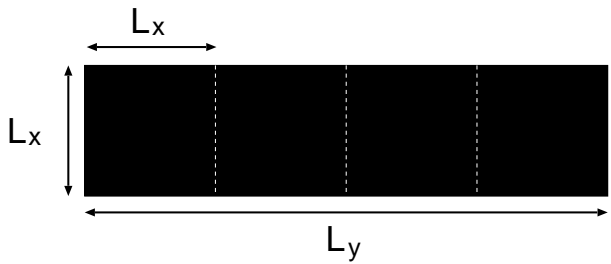}
\caption{\label{devide.contact}
The rubber-countersurface apparent contact area is rectangular
$L_x\times L_y$. We ``divide'' it into $N=L_y/L_x$ square areas with
side $L=L_x$ and area $A_0=L^2$. 
}
\end{figure}

\begin{figure}
\includegraphics[width=0.45\textwidth,angle=0]{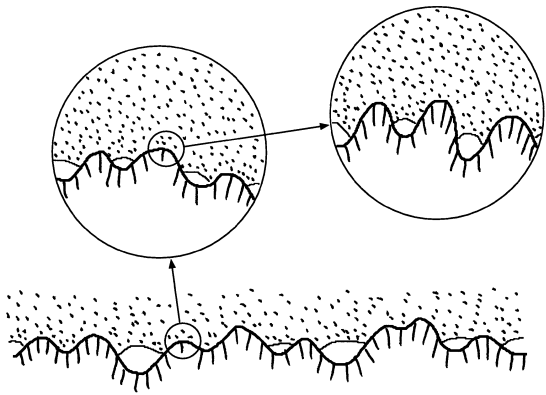}
\caption{\label{1x}
An rubber block (dotted area) in adhesive contact with a hard
rough substrate (dashed area). The substrate has roughness on many different 
length scales and the rubber makes partial contact with the substrate on all length scales. 
When a contact area 
is studied at low magnification it appears as if complete contact occurs, 
but when the magnification is increased it is observed that in reality only partial
contact occurs.  
}
\end{figure}

\begin{figure}
\includegraphics[width=0.45\textwidth,angle=0.0]{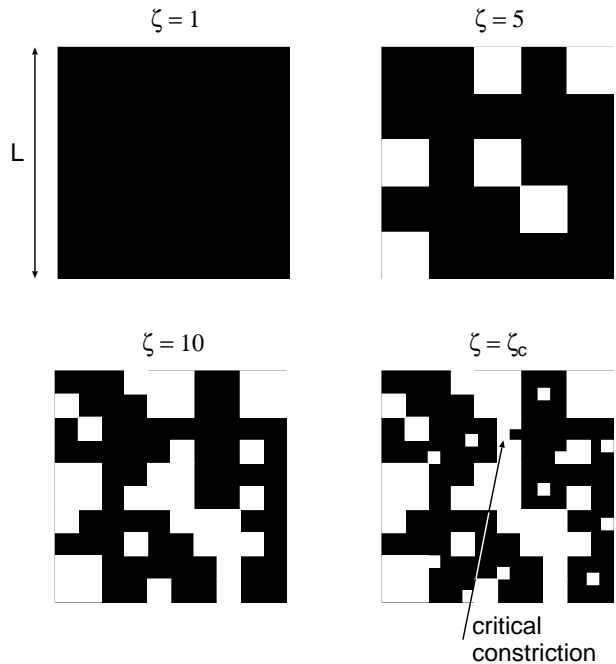}
\caption{\label{percolationpic}
The contact region at different magnifications (schematic). Note that at the
point where the non-contact area (white area) percolate
$A(\zeta_{\rm c}) \approx 0.4 A_0$, while there appear to be complete contact between the
surfaces at the
lowest magnification $\zeta = 1$: $A(1)=A_0$.
}
\end{figure}

\vskip 0.3cm

{\bf 2. Theory}

We first briefly review the basic picture on which our 
calculations of the leak-rate is based\cite{Creton}.
Assume that the nominal contact region between the rubber and the hard countersurface is
rectangular with area $L_x\times L_y$, see Fig. \ref{devide.contact}. 
We assume that the high pressure fluid region is for $x<0$
and the low pressure region for $x>L_x$. We now divide the contact region into squares with
the side $L_x=L$ and the area $A_0=L^2$ (this assumes that $N=L_y/L_x$ is an integer, but this 
restriction does not affect the final result). 
Now, let us study the contact between the two solids within one of the squares
as we change the magnification $\zeta$. We define $\zeta= L/\lambda$, where $\lambda$ is the resolution.
We study how the apparent contact area (projected on the $xy$-plane),
$A(\zeta)$, between the two solids depends on the magnification $\zeta$.
At the lowest magnification we cannot observe any surface roughness, and 
the contact between the solids appears to be complete i.e., $A(1)=A_0$. As we increase the magnification
we will observe some interfacial roughness, and the (apparent) contact area will decrease, 
see Figs. \ref{1x} and \ref{percolationpic}.
At high enough magnification, say $\zeta = \zeta_{\rm c}$, a percolating path of 
non-contact area will be observed 
for the first time, see Fig. \ref{percolationpic}. 
The most narrow constriction along the percolation path will have the lateral
size $\lambda_{\rm c} = L/\zeta_{\rm c}$ and the surface separation at this point is denoted by 
$u_{\rm c}=u_1(\zeta_{\rm c})$
and is given by a recently developed contact mechanics theory (see below). As we continue to
increase the magnification we will find more percolating channels between the surfaces, but these will
have more narrow constrictions 
than the first channel which appears at $\zeta=\zeta_{\rm c}$, and for the moment we will
neglect the contribution to the leak-rate from these channels (see also Sec. 5). 
Thus, in this section we will
assume that the leak-rate is determined by the critical constriction.  

A first rough estimate of the leak-rate is obtained by assuming that all the leakage 
occurs through the critical percolation channel, and that
the whole pressure drop $\Delta P = P_{\rm a}-P_{\rm b}$ (where $P_{\rm a}$ and $P_{\rm b}$ is the 
pressure to the left and right of the
seal) occurs over the critical constriction [of width and length $\lambda_{\rm c} \approx L/\zeta_{\rm c}$
and height $u_{\rm c}={u}_1 (\zeta_{\rm c})$]. Thus for an incompressible
Newtonian fluid, the volume-flow per unit time through the critical constriction
will be 
$$\dot Q = M\Delta P,\eqno(1)$$
where
$$M=\alpha {u_1^3(\zeta_{\rm c}) \over 12 \eta},\eqno(2)$$ 
where $\eta $ is the fluid viscosity. In deriving (1) we have assumed laminar flow and that $u_c << \lambda_c$,
which is always satisfied in practice. 
Here we have introduced a factor $\alpha$ which depends on the
exact shape of the critical constriction, but which is expected to be of order unity. Since there are
$N=L_y/L_x$ square areas in the rubber-countersurface (apparent) contact area, we get the total leak-rate
$$\dot Q = {L_y \over L_x} M\Delta P.\eqno(3)$$ 

To complete the theory we must calculate the separation $u_{\rm c}=u_1(\zeta_{\rm c})$ 
of the surfaces at the
critical constriction. We first determine the critical magnification $\zeta_{\rm c}$ by assuming that the 
apparent relative contact area at this point is given by site percolation theory. 
Thus, the relative contact area $A(\zeta)/A_0 \approx 1-p_{\rm c}$, where $p_{\rm c}$  is the 
so called site percolation threshold\cite{Stauffer}. 
For an infinite-sized systems 
$p_{\rm c}\approx 0.696$ for a hexagonal lattice and $0.593$ for a square lattice\cite{Stauffer}. 
For finite sized systems the percolation will, on the average, occur for (slightly) smaller values
of $p$, and fluctuations in the percolation threshold will occur between 
different realization of the same physical system. We will address this problem again later 
(see Sec. 4) but for
now we take $p_{\rm c}\approx 0.6$ so that $A(\zeta_{\rm c})/A_0 \approx 0.4$ will determine the critical
magnification $\zeta=\zeta_{\rm c}$. 

The (apparent) relative contact area $A(\zeta)/A_0$ at the magnification $\zeta$
can be obtained using the contact mechanics 
formalism developed elsewhere\cite{PSSR,YangPersson,P1,Bucher,JCPpers},
where the system is studied at different magnifications $\zeta$, see Fig. \ref{1x}.
We have\cite{JCPpers,PerssonPRL}

$${A(\zeta)\over A_0} = {1\over (\pi G )^{1/2}}\int_0^{P_0} d\sigma \ {\rm e}^{-\sigma^2/4G} 
= {\rm erf} \left ( P_0 \over 2 G^{1/2} \right )$$
where
$$G(\zeta) = {\pi \over 4}\left ({E\over 1-\nu^2}\right )^2 \int_{q_0}^{\zeta q_0} dq q^3 C(q)$$
where the surface roughness power spectrum
$$C(q) = {1\over (2\pi)^2} \int d^2x \langle h({\bf x})h({\bf 0})\rangle {\rm e}^{-i{\bf q}\cdot {\bf x}}$$
where $\langle ... \rangle$ stands for ensemble average. 
Here $E$ and $\nu$ are the Young's elastic modulus and the Poisson 
ratio of the rubber. The height profile $h({\bf x})$ of the rough surface can be measured routinely
today on all relevant length scales using optical and stylus experiments.

\begin{figure}
\includegraphics[width=0.35\textwidth,angle=0.0]{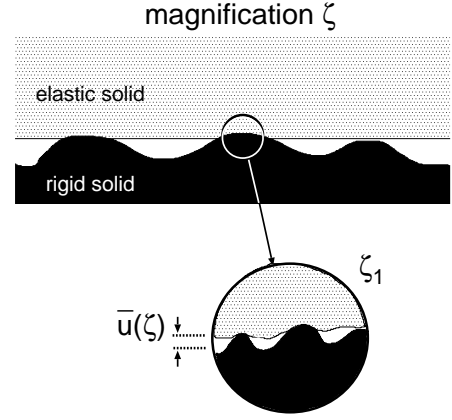}
\caption{\label{asperity.mag}
An asperity contact region observed at the magnification $\zeta$. It appears that
complete contact occur in the asperity contact region, but upon increasing the magnification
it is observed that the solids are separated by the average distance $\bar{u}(\zeta)$.
}
\end{figure}

We define
$u_1(\zeta)$ to be the (average) height separating the surfaces which appear to come into 
contact when the magnification decreases from $\zeta$ to $\zeta-\Delta \zeta$, where $\Delta \zeta$
is a small (infinitesimal) change in the magnification. $u_1(\zeta)$ is a monotonically decreasing
function of $\zeta$, and can be calculated from the average interfacial separation
$\bar u(\zeta)$ and $A(\zeta)$ using
(see Ref. \cite{YangPersson})
$$u_1(\zeta)=\bar u(\zeta)+\bar u'(\zeta) A(\zeta)/A'(\zeta).$$
The quantity $\bar u(\zeta)$ is the average separation between the surfaces in the apparent contact regions
observed at the magnification $\zeta$, see Fig. \ref{asperity.mag}. It can be calculated from\cite{YangPersson}
$$\bar{u}(\zeta ) = \surd \pi \int_{\zeta q_0}^{q_1} dq \ q^2C(q) 
w(q)$$
$$\times \int_{p(\zeta)}^\infty dp' 
 \ {1 \over p'} 
\left [\gamma+3(1-\gamma)P^2(q,p',\zeta)\right ] e^{-[w(q,\zeta) p'/E^*]^2},$$
where $\gamma \approx 0.4$ and where
$$p(\zeta)=P_0A_0/A(\zeta)$$
and
$$w(q,\zeta)=\left (\pi \int_{\zeta q_0}^q dq' \ q'^3 C(q') \right )^{-1/2}.$$
The function $P(q,p,\zeta)$ is given by
$$P(q,p,\zeta) = {2\over \surd \pi} \int_0^{s(q,\zeta)p} dx \ e^{-x^2},$$
where $s(q,\zeta)=w(q,\zeta)/E^*$.

We study the contact between the solids at increasing magnification. In an apparent contact area
observed at the magnification $\zeta$, the substrate has the 
root mean square roughness amplitude\cite{JCPpers,Creton}
$$h_{\rm rms}^2 (\zeta) = 2\pi \int_{\zeta q_0}^{q_1} dq \ q C(q).\eqno(4)$$
When we study the apparent contact area at increasing magnification, the contact pressure
$p(\zeta)$ will increase and the surface roughness amplitude $h_{\rm rms} (\zeta)$ will decrease. Thus,
the average separation $\bar{u}(\zeta)$, between the surfaces in the (apparent) contact regions observed at
the magnification $\zeta$, will decrease with increasing magnification.

\vskip 0.3cm

{\bf 3. Numerical results}

We now present numerical results to illustrate the theory developed above. We assume a rubber
block with a flat surface, squeezed by the nominal pressure $P_0$ against a hard solid with
a randomly rough surface which we assume to be self affine fractal. Thus the surface roughness
power spectrum for $q_0 < q < q_1$:
$$C(q) = C_0 q^{-2(1+H)}$$
where
$$C_0 ={H\over \pi}\langle h^2\rangle \left [q_0^{-2H}-q_1^{-2H}\right ]^{-1}\approx {H\over \pi}
\langle h^2\rangle q_0^{-2H}$$
where $q_0$ and $q_1$ are the long-distance and short-distance cut-off wavevectors, respectively.

The rubber has the Young's modulus $E=10 \ {\rm MPa}$ (as is typical for the low-frequency
modulus of rubber used for seals) and Poisson ratio $\nu = 0.5$. The pressure difference
in the fluid between the two sides of the seals is assumed to be $\Delta P = 0.01 \ {\rm MPa}$.
The fluid is assumed to be an incompressible
Newtonian fluid with the viscosity $\eta = 0.001 \ {\rm Ns/m^2}$. We will study
how the lateral size $\lambda_{\rm c}$ and the height $u_{\rm c}$ of the critical constriction 
depends on the fractal dimension $D_{\rm f}=3-H$ and on the 
root-mean-square roughness amplitude $h_{\rm rms}$ of the rough surface. 
We also present results for how
the volume flow of fluid through the seals depends on $D_{\rm f}$ and $h_{\rm rms}$.
The randomly rough surfaces have the  
cut-off wavevectors $q_0 = 1.0\times 10^4 \ {\rm m}^{-1}$
and $q_1 = 7.8\times 10^9 \ {\rm m}^{-1}$,
and we vary the applied squeezing pressure $P_0$ from $0.05 \ {\rm MPa}$ to $1 \ {\rm MPa}$. 

\begin{figure}
\includegraphics[width=0.45\textwidth,angle=0.0]{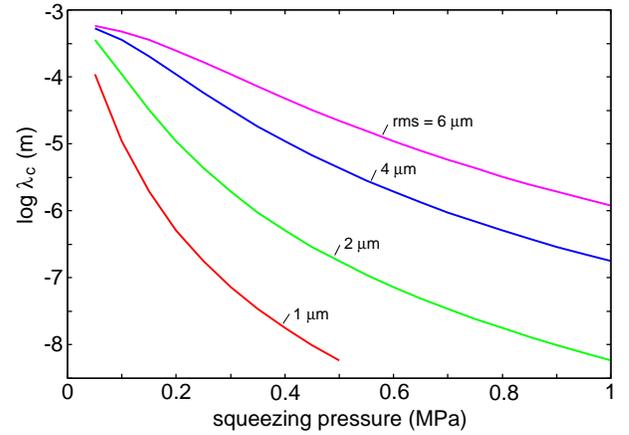}
\caption{\label{p.xlength.rms=1.2.4.6mum.H=0.8}
The lateral size $\lambda_{\rm c}=\lambda (\zeta_{\rm c})$ of the 
critical constriction of the percolation channel,
as a function of the applied normal (or squeezing) pressure $P_0$. Results are shown
for self affine fractal surfaces with the Hurst exponent $H=0.8$ 
(or fractal dimension $D_{\rm f}= 2.2$), and for surfaces
with the root-mean-square roughness (rms) $1,2,4$ and $6\ {\rm \mu m}$.
}
\end{figure}

\begin{figure}
\includegraphics[width=0.45\textwidth,angle=0.0]{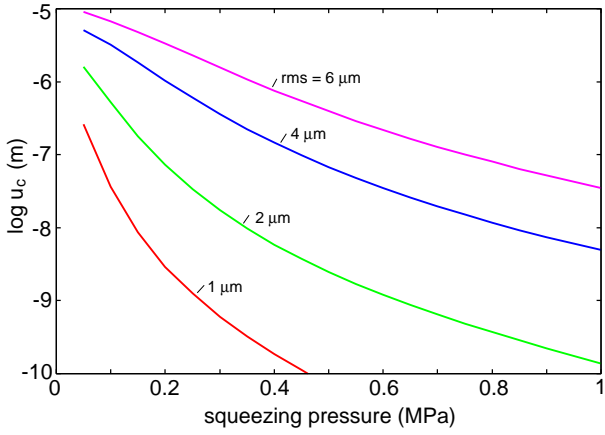}
\caption{\label{p.u.rms=1.2.4.6mum.H=0.8}
The interfacial separation $u_{\rm c}=u_1(\zeta_{\rm c})$ 
at the critical constriction of the percolation channel,
as a function of the applied normal (or squeezing) pressure $P_0$. Results are shown
for self affine fractal surfaces with the Hurst exponent $H=0.8$ 
(or fractal dimension $D_{\rm f}= 2.2$), and for surfaces
with the root-mean-square roughness (rms) $1,2,4$ and $6\ {\rm \mu m}$.
}
\end{figure}

\begin{figure}
\includegraphics[width=0.45\textwidth,angle=0.0]{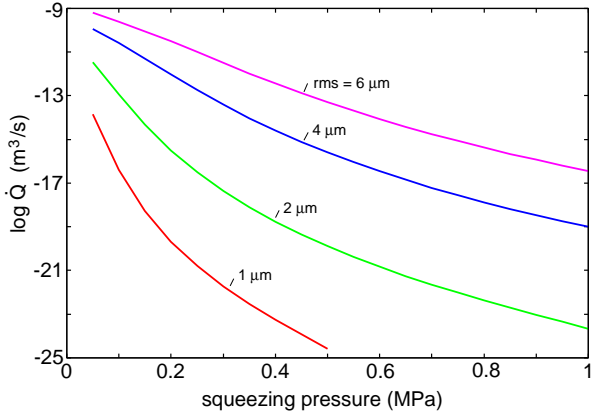}
\caption{\label{p.dotQ.rms=1.2.4.6.mum.H=0.8}
The volume per unit time, $\dot Q$, of fluid leaking through the seals 
as a function of the applied normal (or squeezing) pressure $P_0$. Results are shown
for self affine fractal surfaces with the Hurst exponent $H=0.8$ 
(or fractal dimension $D_{\rm f}= 2.2$), and for surfaces
with the root-mean-square roughness (rms) $1,2,4$ and $6\ {\rm \mu m}$. The fluid pressure
difference between the two sides is $\Delta P = 0.01 \ {\rm MPa}$ and the fluid
viscosity $\mu = 10^{-3} \ {\rm Ns/m^2}$ (water).
}
\end{figure}

Let us first vary the rms roughness amplitude. In Fig. \ref{p.xlength.rms=1.2.4.6mum.H=0.8}
we show the lateral size $\lambda_{\rm c}=\lambda (\zeta_{\rm c})$ 
and in Fig. \ref{p.u.rms=1.2.4.6mum.H=0.8} the height (interfacial separation)
$u_{\rm c}$ of the critical constriction,
as a function of the applied normal (or squeezing) pressure $P_0$. Results are shown
for self affine fractal surfaces with the Hurst exponent $H=0.8$ 
(or fractal dimension $D_{\rm f}= 2.2$), and for surfaces
with the root-mean-square roughness (rms) $1,2,4$ and $6\ {\rm \mu m}$.
As expected, the size of the critical constriction increases when the roughness increases.
In Fig. \ref{p.dotQ.rms=1.2.4.6.mum.H=0.8} we show
the volume per unit time of fluid leaking through the seals 
as a function of the applied normal (or squeezing) pressure $P_0$. 
Note the extremely strong decrease in $\dot Q$ with increasing squeezing pressure and also
its strong dependence on the rms roughness amplitude.

\begin{figure}
\includegraphics[width=0.45\textwidth,angle=0.0]{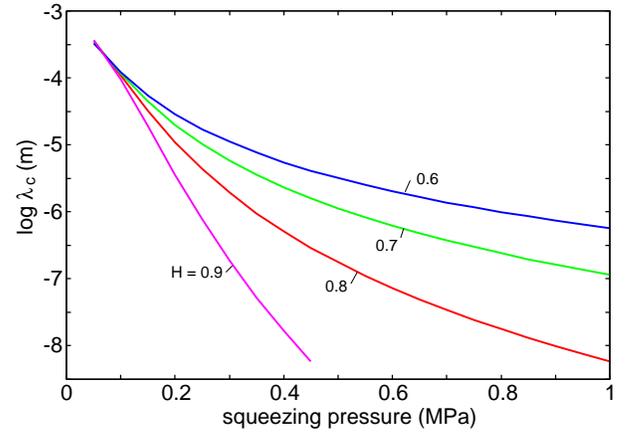}
\caption{\label{p.xlength.rms=2mum.H=0.6.0.7.0.8.0.9}
The lateral size $\lambda_{\rm c}=\lambda (\zeta_{\rm c})$ of the 
critical constriction of the percolation channel,
as a function of the applied normal (or squeezing) pressure $P_0$. 
Results are shown
for self affine fractal surfaces with the root-mean-square roughness (rms) $2 \ {\rm \mu m}$
and for the Hurst exponent $H=0.9$, 0.8, 0.7 and 0.6.
}
\end{figure}

\begin{figure}
\includegraphics[width=0.45\textwidth,angle=0.0]{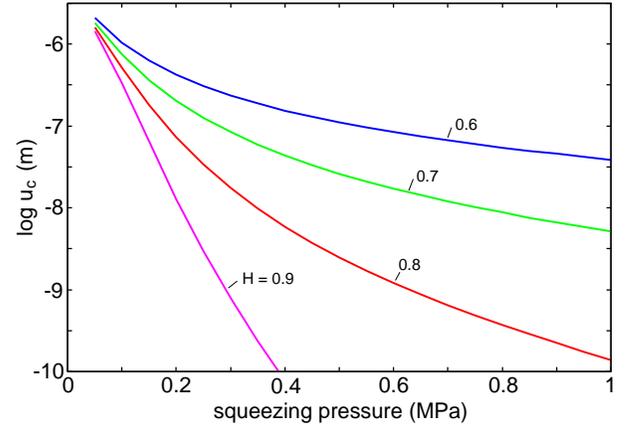}
\caption{\label{p.u.rms=2mum.H=0.6.0.7.0.8.0.9}
The interfacial separation $u_{\rm c}=u_1(\zeta_{\rm c})$ 
at the critical constriction of the percolation channel,
as a function of the applied normal (or squeezing) pressure $P_0$. 
Results are shown
for self affine fractal surfaces with the root-mean-square roughness (rms) $2 \ {\rm \mu m}$
and for the Hurst exponents $H=0.9$, 0.8, 0.7 and 0.6.
}
\end{figure}

\begin{figure}
\includegraphics[width=0.45\textwidth,angle=0.0]{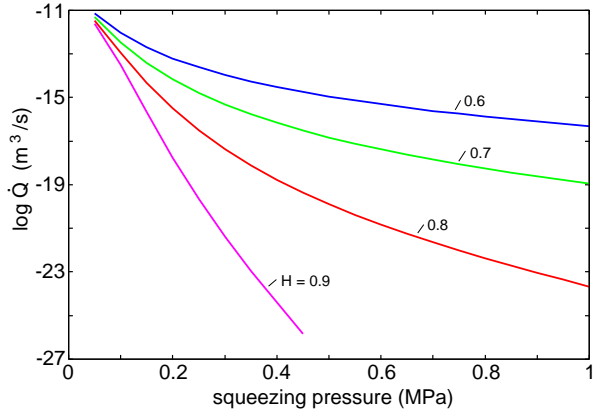}
\caption{\label{p.dotQ.rms=2mum.H=0.6.0.7.0.8.0.9}
The volume per unit time, $\dot Q$, of fluid leaking through the seal 
as a function of the applied normal (or squeezing) pressure $P_0$. 
Results are shown
for self affine fractal surfaces with the root-mean-square roughness (rms) $2 \ {\rm \mu m}$
and for the Hurst exponent $H=0.9$, 0.8, 0.7 and 0.6.
The fluid pressure
difference between the two sides is $\Delta P = 0.01 \ {\rm MPa}$ and the fluid
viscosity $\mu = 10^{-3} \ {\rm Ns/m^2}$ (water).
}
\end{figure}

In Figs. 
\ref{p.xlength.rms=2mum.H=0.6.0.7.0.8.0.9}, \ref{p.u.rms=2mum.H=0.6.0.7.0.8.0.9}
and \ref{p.dotQ.rms=2mum.H=0.6.0.7.0.8.0.9}
we show the analogous results when we vary the Hurst exponent $H=0.9$, 0.8, 0.7 and 0.6
for $h_{\rm rms} = 2 \ {\rm \mu m}$. Note that 
when $H$ decreases for a fixed $h_{\rm rms}$, the short wavelength 
roughness increases while the long wavelength roughness is almost unchanged.

\begin{figure}
\includegraphics[width=0.45\textwidth,angle=0.0]{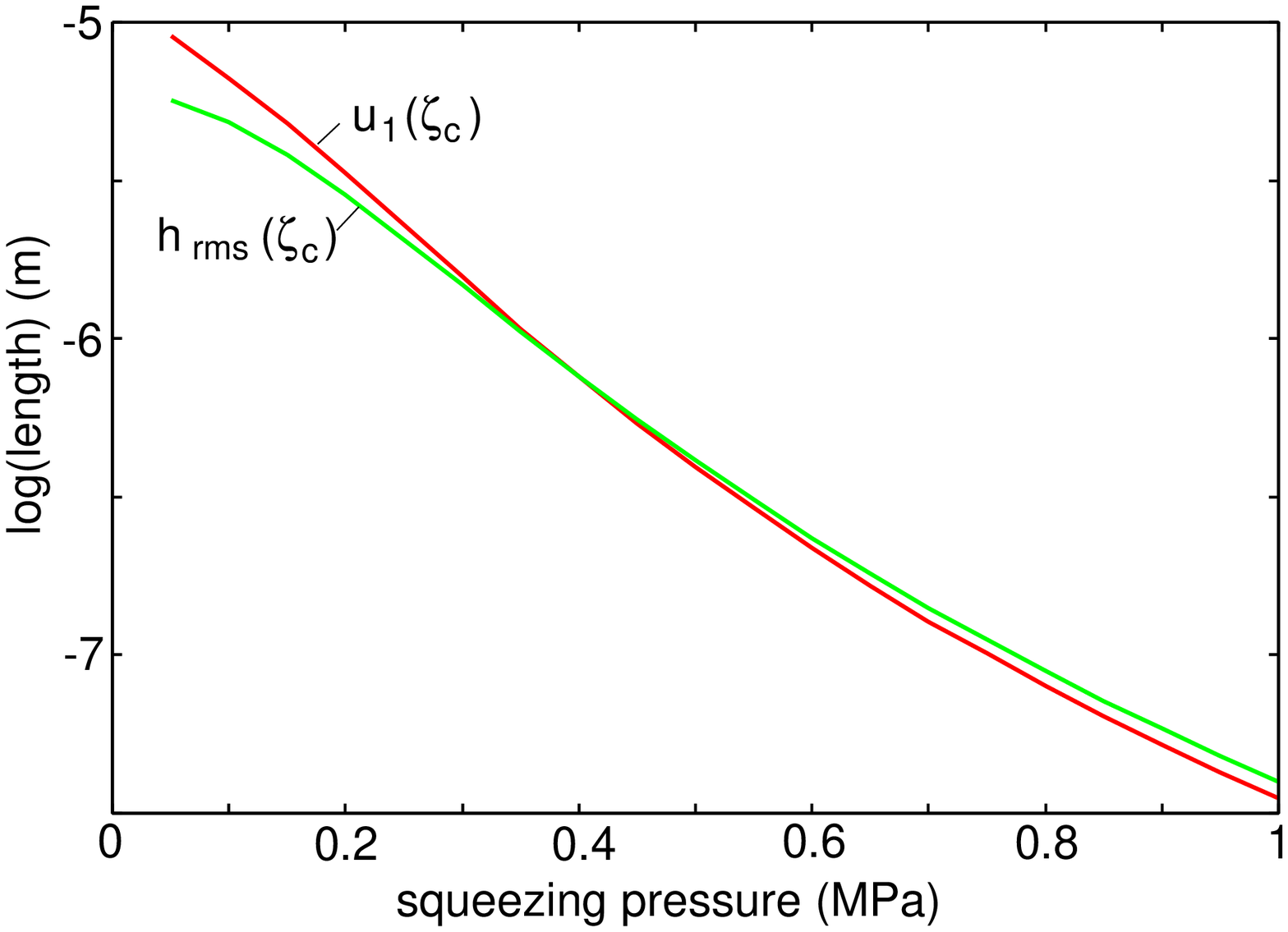}
\caption{\label{p.hrms.uc.H=0.8.rms=6mum}
The interfacial separation $u_{\rm c}=u_1(\zeta_{\rm c})$ and the 
rms-roughness $h_{\rm rms}(\zeta_{\rm c} )$ in the critical constriction of the percolation channel,
as a function of the applied normal (or squeezing) pressure $P_0$. Results are shown
for self a affine fractal surfaces with the Hurst exponent $H=0.8$ 
(or fractal dimension $D_{\rm f}= 2.2$), and 
with the root-mean-square roughness (rms) $6\ {\rm \mu m}$.
}
\end{figure}

In Fig. \ref{p.hrms.uc.H=0.8.rms=6mum}
we show the interfacial separation $u_1(\zeta_{\rm c})$ and the 
rms-roughness $h_{\rm rms}(\zeta_{\rm c} )$ in the critical constriction,
as a function of the applied normal (or squeezing) pressure $P_0$. Results are shown
for a self affine fractal surface with the Hurst exponent $H=0.8$ 
(or fractal dimension $D_{\rm f}= 2.2$), and 
with the root-mean-square roughness (rms) $6\ {\rm \mu m}$. Note that the difference between
$h_{\rm rms}(\zeta_{\rm c} )$ and $u_1(\zeta_{\rm c})$ is relatively small. 
We have found that this is the case
also for the other parameters used in the study above.

\vskip 0.3cm

{\bf 4. Molecular Dynamics results}

\begin{figure}
\includegraphics[width=0.45\textwidth,angle=0.0]{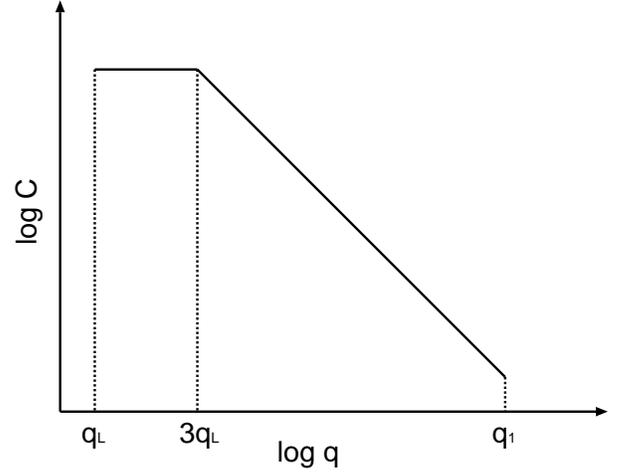}
\caption{\label{logC.logq}
Surface roughness power spectrum of a surface which is self-affine fractal for 
$q_1>q>3q_L$. The slope ${\rm log} C-{\rm log} q$ relation for $q>3q_L$ determines the
fractal exponent of the surface. The lateral size $L$ of the surface determines
the smallest wave-vector $q_L=2\pi/L$.} 
\end{figure}

\begin{figure}
\includegraphics[width=0.45\textwidth,angle=0.0]{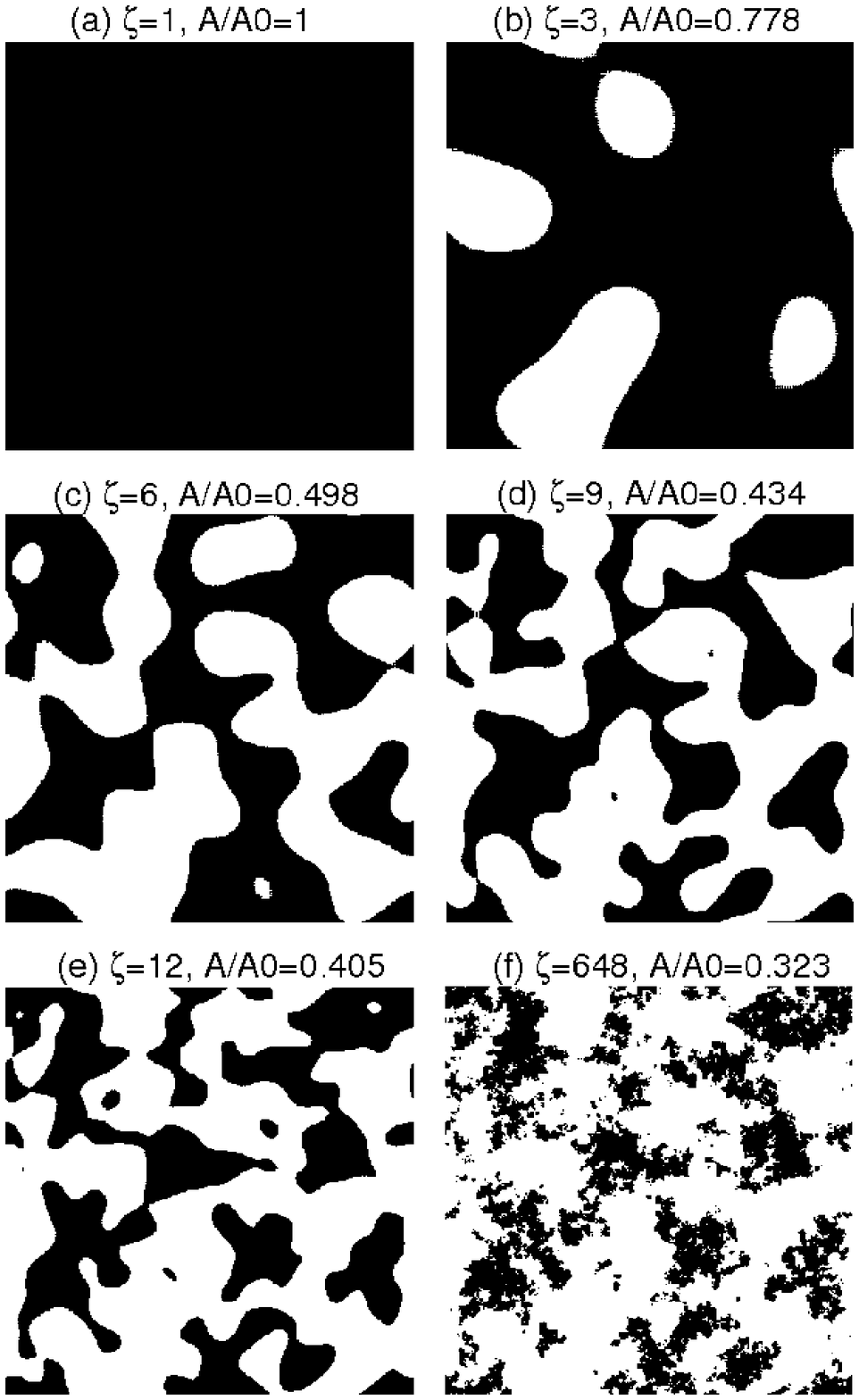}
\caption{\label{sealing.MD}
The contact regions at different magnifications $\zeta=1, 3, 6, 9, 12, 648$,
are shown in (a)-(f) respectively. The pressure is $p \approx 4.1 \ \rm GPa$. When
the magnification is increased from 9 to 12, the non-contact region percolate.}
\end{figure}

The multiscale molecular dynamics model has been described in Ref.\cite{Chunyan},
but we review it briefly here. In what follows we denote the lower solid
as {\it substrate}, the upper solid as {\it block}.
We are concerned with the contact between 
a randomly rough and rigid substrate, and an elastic block, without adhesion.
We are interested in surfaces with random roughness with wave-vector 
components in the finite range $q_1>q>q_L$ (see Fig. \ref{logC.logq}), 
where $q_L=2\pi/L$, $L$ is the lateral size of the system. 
In order to accurately study contact mechanics between elastic solids,
it is necessary to consider a solid block which extends a distance $\sim L$
in the direction normal to the nominal contact area. This requires huge number
of atoms or dynamical variables even for small systems. Therefore, we developed
a multiscale molecular dynamics approach to study contact mechanics, to avoid 
this trouble\cite{Chunyan}. The lateral size of the system is $L=1040 \ \rm \AA$.
$L_x=N_xa$ and $L_y=N_ya$, where $a=2.6 \ \rm \AA$ is the lattice space of the 
block, $N_x=N_y=400$ for the block. The elastic modulus and Poisson ratio
are $E=77.2 \ \rm GPa$ and $\nu=0.42$. The lattice space of the substrate is
$b \approx a/\phi$, where $\phi = \left(1+\sqrt5\right)/2$ is the golden mean, in
order to achieve (nearly) incommensurate structures at the interface.

For self-affine fractal surfaces, the power spectrum has power-law behavior
$C(q) \sim q^{-2(H+1)}$, where the Hurst exponent $H$ is related
to fractal dimension $D_f$ of the surface via $ H=3-D_f$.
For real surfaces this relation holds only for a finite wave vector 
region $q_1<q<q_0$, where $q_1=2\pi/b$, $q_0$ is roll-off wave-vector
$q_0=3q_L$ (see Fig. \ref{logC.logq}). The randomly rough surfaces have been
generated as described in Ref. \cite{Chunyan,P3}, which have root-mean-square
roughness $h_{rms}=10 \ \rm \AA$ and fractal dimension $D_f=2.2$.
The roll-off wave-vector $q_0=3q_L$, where $q_L=2\pi/L$ and $L=1040 \ \rm \AA$.
In this section we define the magnification $\zeta=q/q_L$.

The atoms at the interface between block and substrate interact with repulsive
potential $U(r)=\epsilon(r_0/r)^{12}$, where $r$ is the distance between 
a pair of atoms, $r_0=3.28 \ \rm \AA$ and $\epsilon=74.4 \ \rm meV$.
In molecular dynamics simulations there is no unique definition of contact
(see Ref. \cite{Chunyan}). Here we use the critical distance $d_c$ to define 
contact. If the separation between two atoms is smaller than $d_c$, it has been
denoted as contact, otherwise non-contact. Here $d_c=4.36 \ \rm \AA$.
  
Fig. \ref{sealing.MD}
shows the block-substrate contact region at different magnifications $\zeta=1, 3, 6, 9, 12, 648$.
Note that when the magnification is increased from 9 to 12, the non-contact region 
percolates.
The percolation occurs when the normalized projected contact area $A/A_0 \approx 0.4$, in good 
agreement with percolation theory\cite{Stauffer}.

\vskip 0.3cm

{\bf 5. Improved analytical description}

\begin{figure}
\includegraphics[width=0.46\textwidth,angle=0.0]{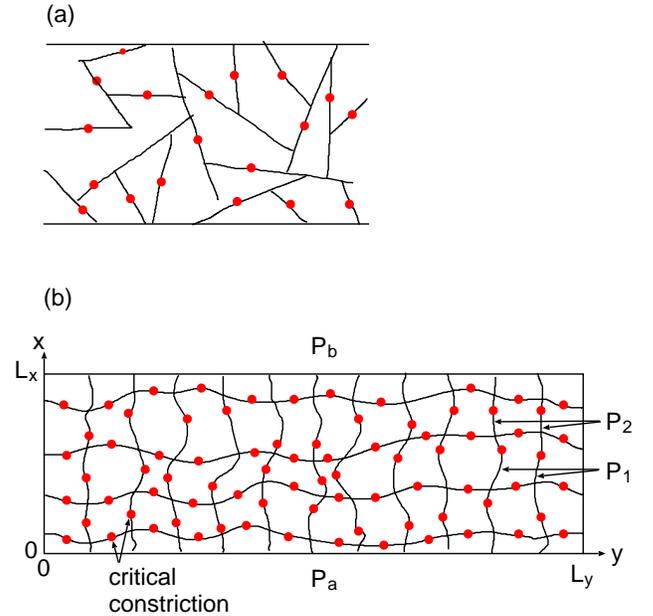}
\caption{\label{many.junctions}
The solid lines denote non-contact channels and the dots critical or near critical
constrictions. In reality the constrictions and channels are nearly randomly distributed as
in (a) [see also Fig. \ref{sealing.MD}(e)] but in the model 
calculation we use the more ordered structure shown in (b).
}
\end{figure}

In Sec. 2 we assumed that all the fluid flow occurs through a single 
constriction which we refer to as the critical constriction. In reality, fluid flow will
also occur in other flow channels even if they have more narrow constrictions. In this section
we will assume that there are a finite concentration of critical or near critical constrictions, which
correspond to all constrictions which appear when the magnification changes in some narrow interval
around the critical value $\zeta_{\rm c}$, 
e.g., in such a way that $A(\zeta)/A_0$
changes by, say, $\pm 0.03$. Since we are very close to the percolation threshold, we will assume
that the size of all the (nearly critical) constrictions remains the same. 
In a more accurate treatment one would instead introduce a distribution of sizes of constrictions.
In Fig. \ref{many.junctions}(a) the dots correspond to
the critical or near critical constrictions along percolation channels (solid lines). 
One expects the (nearly critical) constrictions to be nearly randomly distributed
in the apparent contact area, and that the channels, of which they are part, to have all possible
directions as indicated by the lines in Fig. \ref{many.junctions}(a). Here we will consider
a simplified version of (a) where the (nearly critical) constrictions form a more ordered arrangement
as in Fig. \ref{many.junctions}(b).
In reality, the dots and the lines should be (nearly) 
randomly distributed as in \ref{many.junctions}(a), but this is likely to have only
minor effects on what follows. 

On the average fluid will only flow in the $x$-direction. Thus, in a first approximation
one may assume that no fluid flows along the (transverse)  
channels pointing (mainly) in the $y$-direction
in Fig. \ref{many.junctions}(b). Let $a$ be the (average) distance between two nearby critical
constrictions. Thus we expect $n=L_x/a$ constrictions along a 
percolation channel (in  the figure we have $n=3$).
If $\dot Q_1$ denotes the fluid
volume per unit time flowing along one percolation channel, then we must have
$$\dot Q_1= M (P_{\rm a}-P_1) = M(P_1-P_2)=
...=M(P_n-P_{\rm b}).\eqno(5)$$
From (5) we get
$$\dot Q_1= {M\over n} (P_{\rm a}-P_{\rm b})$$
As expected, the amount of fluid flowing in the channel is reduced when the number
of constriction increases. However, there will be roughly $L_y/a$ percolation channels so the total 
fluid flow will be
$$\dot Q= {L_y\over L_x}M (P_{\rm a}-P_{\rm b})\eqno(6)$$
which is identical to the result obtained in Sec. 2. This analysis is very rough, and a more detailed
analysis will result in some modifications of the leak-rate, but (6) should be very useful as 
a first rough estimate of the leak-rate.
Note that the present treatment will result in a more gradual changes in the liquid
pressure in the apparent contact region, from the initial high pressure value $P_{\rm a}$ (entrance side) 
to the low pressure value $P_{\rm b}$ (exit side).

\begin{figure}
\includegraphics[width=0.46\textwidth,angle=0.0]{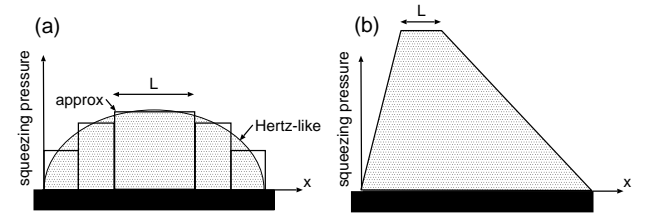}
\caption{\label{pressure.lip.Oring}
Nominal contact pressure distribution $P_0(x)$ (curve bounding the dotted area) for (a) O-ring seal and (b)
lip seal. In (a) the curve denoted by ``approx'' is an approximation to the continues
Hertz-like curve.  
}
\end{figure}

\vskip 0.3cm

{\bf 6. Comparison with experiment}

We have not found any results in the literature about leak-rates of seals for well characterized systems.
However, we have found some results which are in qualitative agreement with our theory. 
For example, leak-rates
observed for both rubber and steel seals tend to decrease very fast (roughly exponentially) 
with increasing contact force. 
Thus, in Ref. \cite{Widder} the leak-rate for a rubber seal decreased 
by 6 order of magnitude as the load increased
by a factor of 10. A similar sharp drop in the leak-rate with increasing contact force 
has been observed for seals made from steel\cite{Schmidt}. 
However, in the latter case some plastic deformation is likely
to occur in the contact region. In both cases the nominal pressure may change less 
than the change in the load,
due to an increase in the nominal contact area with increasing load. A detailed analysis
of the experimental data is not possible as the surface topography was not studied in detail.
In Sec. 9 we suggest a simple experiment which can be used to test the theory. 

The present theory implys that most of the fluid leakage occurs through the critical or nearly
critical constrictions in the percolating channels at the interface between the two solids.
Since the constrictions are very small they can easily be clogged up by dirt particles in the 
fluid. This results in leak-rates which decreases with increasing time as the microscopic gaps
get clogged up. This has recently been observed for metal seals\cite{Schmidt}. In fact, by
using specially prepared fluids with immersed particles having a narrow distribution of particle 
diameters, it should be possible to determine (approximately) the size (or rather the height)
of the critical constriction.

\vskip 0.3cm

{\bf 7. Comment on the role of non-uniform pressure, rubber viscoelasticity and adhesion}

In the study above we have assumed that the normal (squeezing) pressure is constant in the 
nominal rubber-countersurface contact region. In reality, this is (almost) never the case. Thus, in
rubber O-ring applications one expect a pressure distribution which is Herzian-like, as indicated in Fig. 
\ref{pressure.lip.Oring}(a). In (dynamical) 
rubber seals for linear reciprocal motion, the 
pressure distribution is asymmetric, with a much steeper increase in the pressure when
going from the high-pressure ($P_{\rm a}$) fluid side towards 
the center of the seal, as compared going from the 
low-pressure ($P_{\rm b}$) fluid side toward the center of the seal, see Fig. \ref{pressure.lip.Oring}(b).
(The reason for this asymmetry does not interest us here.) 
The theory developed above can be applied approximately to these cases too. Thus in case (a) 
(e.g., rubber O-ring seals) one may approximate the actual Herzian-like pressure profile with a sum of
step functions as indicated in Fig. \ref{pressure.lip.Oring}(a). Since the seal-action is so strongly dependent
on the squeezing pressure (see Figs. \ref{p.dotQ.rms=1.2.4.6.mum.H=0.8} 
and \ref{p.dotQ.rms=2mum.H=0.6.0.7.0.8.0.9}), 
it  is enough to include the central step region (with width $L$) in the analysis.
Since there is no unique way to determine the width $L$ there will be some (small) uncertainty in the
analysis, but this is not important in most practical cases. Similarly, for the lip-seal 
[Fig. \ref{pressure.lip.Oring}(b)] during stationary condition it is enough to include the 
region (width $L$) where the normal pressure is maximal.

In the study above we have assumed that the rubber behaves as a purely elastic solid. In reality,
rubber-materials are viscoelastic. One consequence of this is stress relaxation. For example, after
a rubber O-ring has been deformed to fit into the ``cavity'' where it is placed, the stress exerted
on the solid walls will decrease with increasing time. Since rubber materials have very wide distribution
of relaxation times, the stress can continue to 
decrease even one year after installation. Thus, after very long time the pressure
in the rubber-countersurface contact region may be so low that the seal fails (note: we found earlier
that the leak-rate depends extremely sensitively on the normal pressure). Stress relaxation can be easily taken
into account in the analysis above by using the relaxation modulus $E(t)$ (where $t$ is time) measured
in the laboratory using standard methods.

Finally, let us comment on the role of adhesion in rubber seals. We first note that if the fluid
is an oil, the effective adhesion between the rubber and the hard countersurface may vanish
or nearly vanish, as observed in some experiments\cite{Zappone}. If the fluid is not an oil (e.g., water)
some effective adhesive interaction may remain. In particular, if the fluid is a gas then the  
the effective adhesion may be similar to that in the normal atmosphere. However, even in this case the
adhesive interaction between the solids may have a negligible influence on the leak-rate. The reason for this
is that adhesion operates mainly at very short length scales, corresponding to 
high magnification $\zeta > \zeta_{\rm ad}$, 
while the leak-rate is determined mainly by the contact mechanics at the point where the first percolation
channel appears, corresponding to the magnification $\zeta_{\rm c}$. 
If $\zeta_{\rm c} << \zeta_{\rm ad}$
the adhesive interaction will have a negligible influence on the leak-rate. We now illustrate this with
a numerical calculation using the theory of Ref. \cite{P1}.

\begin{figure}
\includegraphics[width=0.46\textwidth,angle=0.0]{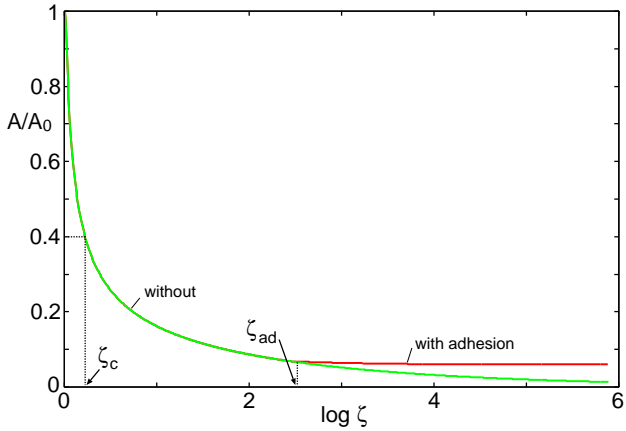}
\caption{\label{log.zeta.Area.hrms=6mum.H=0.8}
The relative contact area $A(\zeta)/A_0$ as a function of the logarithm 
of the magnification $\zeta$. Note that at the magnification $\zeta_{\rm c}$, where 
the non-contact area first percolate, the adhesional interaction has no influence on the
contact area. The adhesional interaction will manifest itself only for
$\zeta > \zeta_{\rm ad}$, where the adhesional interaction increases the contact area as compared to the
case without the adhesional interaction included.
For rubber block in contact with a hard solid with a self-affine fractal surface 
with the root-mean-square roughness $h_{\rm rms} = 6 \ {\rm \mu m}$,
the Hurst exponent $H=0.8$. The squeezing pressure $P_0=0.2 \ {\rm MPa}$ and, 
for the curve ``with adhesion'', 
with the interfacial binding energy
per unit area $\Delta \gamma = 0.05 \ {\rm J/m^2}$.
}
\end{figure}

Fig. \ref{log.zeta.Area.hrms=6mum.H=0.8} shows 
the relative contact area $A(\zeta)/A_0$ as a function of the logarithm 
of the magnification $\zeta$. Note that at the magnification $\zeta_{\rm c}$, where 
the non-contact area first percolates, the adhesional interaction has no influence on the
contact area. The adhesional interaction will manifest itself only for
$\zeta > \zeta_{\rm ad}$, where the adhesional interaction increases the contact area as compared to the
case without the adhesional interaction included.
The results in Fig. \ref{log.zeta.Area.hrms=6mum.H=0.8} 
is for rubber block in contact with a hard solid with a self-affine fractal surface 
with the root-mean-square roughness $h_{\rm rms} = 6 \ {\rm \mu m}$,
the Hurst exponent $H=0.8$, and for the squeezing pressure $P_0=0.2 \ {\rm MPa}$.

Finally we note that if there is very little fluid at the interface strong capillary adhesion may occur 
between the surfaces. This is known to be of great importance in, e.g., 
the context of rubber wiper blades. This topic has been discussed in detail in Ref. \cite{Per,Del}

\vskip 0.3cm

{\bf 8. Dynamical seals}

\begin{figure}
\includegraphics[width=0.40\textwidth,angle=0.0]{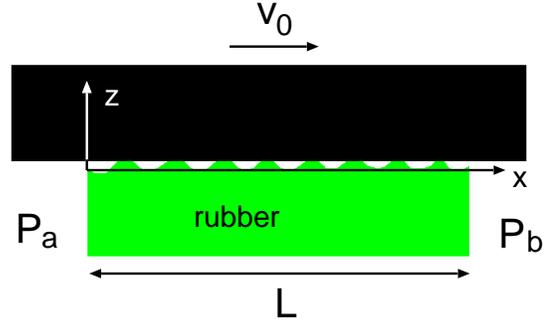}
\caption{\label{rubber.hard.counter}
A rubber block with a rough surface in contact with a hard smooth countersurface (upper block)
which moves relative to the rubber block with the velocity $v_0$. 
}
\end{figure}

The theory presented above is for static seals. Here we give some comments related to 
dynamical seals. We will estimate the leak-rate for linear reciprocal seals at very low
sliding velocity. We assume that the roughness occurs mainly on the rubber surface and we treat 
the hard countersurface as perfectly flat. Thus, as the rubber slides along the countersurface
the contact mechanics does not change, e.g., the percolation channel will be time-independent in the
reference frame moving with the rubber. We consider the system in the reference frame where the rubber
is stationary while the hard countersurface moves from left to right with the velocity $v_0$.
The rubber is assumed to be below the countersurface, see Fig. \ref{rubber.hard.counter}. 
The high pressure fluid 
region (pressure $P_{\rm a}$) occupys $x < 0$ 
while the low pressure region (pressure $P_{\rm b}$) occupys $x> L$.

We assume a Newtonian fluid and stationary and laminar flow. The basic equations for the fluid flow are
$$\nabla p = \eta \nabla^2 {\bf v}, \ \ \ \ \ \ \ \ \nabla \cdot {\bf v} = 0,$$
where $p({\bf x})$ and ${\bf v} ({\bf x})$ are the fluid pressure and the fluid flow velocity, respectively.
We now consider the fluid flows in the percolation channel. 
Let $s$ be the length-coordinate along the percolation channel. Since in general the $\lambda (s) >>
u(s)$, where $\lambda (s)$ is the width and $u(s)$ the height of the channel at the point $s$ along
the channel, we can write the velocity as ${\bf v}({\bf x}) = \hat s v(s,z)$ where
$$v(s,z) \approx {1\over 2 \eta} {dp \over ds} z(z-u(s))+v_0 \hat x \cdot \hat s {z\over u(s)}$$
The volume flow per unit time through any cross section of the channel is assumed to be the same, 
and equal to $\dot Q$ which gives
$$\dot Q = \lambda (s)\int_0^{u(s)} dz \ v(s) = $$
$$\lambda (s)\left (-{u^3(s)\over 12 \eta} {dp \over ds}
+v_0 \hat x \cdot \hat s {u(s)\over 2}\right )$$ 
or
$${dp \over ds} = {6 \eta \over u^2(s)}v_0 \hat x \cdot \hat s- {12 \eta \dot Q \over \lambda(s) u^3(s)}$$
Integrating this equation gives
$$p(l_{\rm a})=P_{\rm a}+
\int_0^{l_{\rm a}} ds \ \left [{6 \eta \over u^2(s)}v_0 \hat x \cdot \hat s- 
{12 \eta \dot Q \over \lambda(s) u^3(s)} \right ]\eqno(7)$$ 
where $l_{\rm a}$ is the length of the percolation channel.
Let $\tilde P(u)$ be the probability to find the surfaces separated by the hight $u$ along the
percolation channel. Note that $\lambda(s)$ also can (at least locally) be considered as a function
of $u$, which we denote by $\lambda (u)$ for simplicity. Thus we can write (7) as
$$p(l_{\rm a})=P_{\rm a}+
6 L_{\rm a} \eta v_0 \int_{u_{\rm c}}^\infty du \ {\tilde P(u) \over u^2 } 
- 12 l_{\rm a} \eta \dot Q \int_{u_{\rm c}}^\infty du \ {\tilde P(u) \over \lambda(u) u^3}\eqno(8)$$ 
where $L_{\rm a}$ is the length of the percolation path projected on the $x$-axis.
In Ref. \cite{YangPersson} we have shown how it is possible to calculate the distribution $\bar P_u$
of heights $u$ between two surfaces in elastic contact. We will now assume that (note: $u> u_{\rm c}$) 
$$\tilde P(u) \approx {{\bar P_u} \over \int_{u_{\rm c}}^\infty du' \ {\bar P_{u'}}}$$ 
Let us write (8) as
$$p(l_{\rm a})=P'_{\rm a} = P_{\rm a}+B_{\rm a}v_0-C_{\rm a}\dot Q\eqno(9)$$
where
$$B_{\rm a}= 6 L_{\rm a} \eta \int_{u_{\rm c}}^\infty du \ {\tilde P(u) \over u^2 }\eqno(10)$$
and
$$C_{\rm a}= 12 l_{\rm a} \eta \int_{u_{\rm c}}^\infty du \ {\tilde P(u) \over \lambda(u) u^3}\eqno(11)$$
Similarly, one gets\cite{comment1} 
$$p(l_{\rm b})=P'_{\rm b} = P_{\rm b}-B_{\rm b}v_0-C_{\rm b}\dot Q\eqno(12)$$
Thus in this case (1) takes the form
$$\dot Q = M(P'_{\rm a}-P'_{\rm b})= $$
$$M(P_{\rm a}-P_{\rm b})+M(B_{\rm a}+B_{\rm b})v_0-
M(C_{\rm a}-C_{\rm b})\dot Q$$
or
$$\dot Q= M {\Delta P +(B_{\rm a}+B_{\rm b})v_0 \over 1- M(C_{\rm a}-C_{\rm b})}\eqno(13)$$
The factor $ M(C_{\rm a}-C_{\rm b})$ in the
denominator in this expression is independent of $v_0$ and we will assume 
that it is negligible compared to unity, and neglect it. One interesting application of
(13) is to wiper blade. Here $\Delta P = 0$ so that (13) takes the form
$$\dot Q = 
M(B_{\rm a}+B_{\rm b})v_0\eqno(14)$$
Substituting (2) and (10) (and a similar expression for $B_{\rm b}$) in (14) gives
$$\dot Q =L_y u_{\rm c}^3 v_0 {\alpha \over 2} \int_{u_{\rm c}}^\infty du \ {\tilde P(u) \over u^2}\eqno(15)$$
where we have included the extra factor $L_y/L_x$ to take into account the number of square seal units.
During the time $t$ the leak-volume is $\dot Q t$. We define the average thickness $d$ of the leak-film
as $d=\dot Q t/(L_y v_0t)$. From (15) we get
$$d=\beta u_{\rm c}\eqno(16)$$
$$\beta = {\alpha \over 2} \int_{u_{\rm c}}^\infty du \ u_{\rm c}^2 {\tilde P(u) \over u^2}\eqno(17)$$ 
We have calculated the integral $I$ in $\beta$ for some typical cases\cite{example},
and found that $I \approx 0.1-0.2$ so we expect
$\beta \approx 0.1$.

In the treatment above we have assumed that the
contact between the rubber and the hard countersurface does not
depend on the pressure in the fluid, which is a good approximation
as long as the fluid pressure $p({\bf x}) << P_0$. However, as the sliding
velocity increases, the fluid pressure in some regions at the interface will increase
which will tend to increase the separation between the two surfaces. At very high
sliding velocity, hydrodynamic lubrication will prevail and the surfaces are
completely separated by a thin fluid film. However, even at much lower sliding velocity the
hydrodynamic pressure buildup may strongly increase the leak rate. In particular, the pressure
at the critical constriction will tend to increase the separation between the surfaces
and hence increase the leak-rate. We will not study this effect here but just estimate 
when this effect becomes important. Let $p_{\rm c}$ be the pressure at the critical constriction.
If $p_{\rm c}$ acts over the area $\lambda_c^2$ it will locally increase the separation between
the surfaces by the amount \cite{comment2}
$\Delta u \approx \lambda_{\rm c} p_{\rm c} /E$. Thus, the pressure at the critical constriction
must be much smaller than $E u_{\rm c} /\lambda_{\rm c}$ in order for the pressure-induced 
effect to be negligible. Note that
$$p_{\rm c} \approx 6L \eta v_0 \int_{u_{\rm c}}^\infty du \ {\tilde P(u) \over u^2}$$
so the present study is limited to sliding velocities
$$v_0 << {E u^3_{\rm c} \over 6 L \eta \lambda_{\rm c}} \left (\int_{u_{\rm c}}^\infty 
du \ u^2_{\rm c} {\tilde P(u) \over u^2} \right )^{-1} 
\approx {E u^3_{\rm c} \over L \eta \lambda_{\rm c}}\eqno(17)$$
where we have used that the integral typically is of order $\sim 0.15$.
Thus, for example, if in a wiper blade application\cite{Koenen}, 
after some use the rubber blades typically develop (because of wear) a 
surface roughness with a rms amplitude of several micrometer. If the rms roughness
is $2 \ {\rm \mu m}$ (and the Hurst exponent $H=0.8$), the nominal pressure 
$\sim 0.2 \ {\rm MPa}$, and if we assume that $E\approx 10 \ {\rm MPa}$ 
we get from Fig. \ref{p.u.rms=1.2.4.6mum.H=0.8} and \ref{p.xlength.rms=1.2.4.6mum.H=0.8} 
$u_{\rm c} \approx 0.1 \ {\rm \mu m}$
and $\lambda_{\rm c} \approx 10 \ {\rm \mu m}$. If $L\approx 0.1 \ {\rm mm}$ and (for water) $\eta \approx
10^{-3} \ {\rm Pa s}$ we get that the slip velocity must be at most $\sim 1 \ {\rm cm/s}$ in order for
(17) to be valid. According to (16) the (average) film thickness of the water layer would be of order
$0.01 \ {\rm \mu m}$.   

\vskip 0.3cm

{\bf 9. A new experiment}

\begin{figure}
\includegraphics[width=0.30\textwidth,angle=0.0]{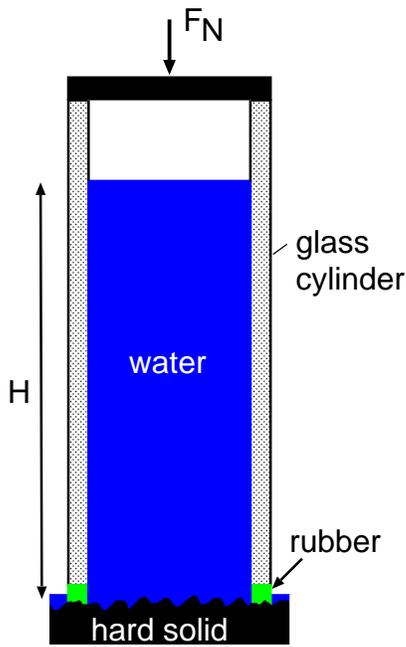}
\caption{\label{tube.water}
A simple experimental set-up for measuring the leak-rate of seals.
A glass (or PMMA) cylinder with a rubber ring glued to one end is squeezed against
a hard substrate with well-defined surface roughness. The cylinder is filled with 
a fluid, e.g., water, and the leak-rate of the fluid at the rubber-countersurface
is detected by the change in the height of the fluid in the cylinder. In this case
the pressure difference $\Delta P = P_{\rm a}-P_{\rm b} = \rho g H$, where $g$ is the gravitation
constant, $\rho$ the fluid density and $H$ the height of the fluid column. With $H\approx 1 \ {\rm m}$
we get typically $\Delta P \approx 0.01 \ {\rm MPa}$.
}
\end{figure}

Very few studies of leak rates of seals with well characterized surfaces
have been published. Here we would like to suggest a very simple 
experiment which could be used to test the theory presented in Sec. 2. 
In Fig. \ref{tube.water} we show a 
set-up for measuring the leak-rate of seals.
A glass (or PMMA) cylinder with a rubber ring (with rectangular cross-section)
glued to one end is squeezed against
a hard substrate with well-defined surface roughness. The cylinder is filled with 
a fluid, e.g., water, and the leak-rate of the fluid at the rubber-countersurface
is detected by the change in the height of the fluid in the cylinder. In this case
the pressure difference $\Delta P = P_{\rm a}-P_{\rm b} = \rho g H$, where $g$ is the gravitation
constant, $\rho$ the fluid density and $H$ the height of the fluid column. With $H\approx 1 \ {\rm m}$
we get typically $\Delta P \approx 0.01 \ {\rm MPa}$. With the diameter of the glass cylinder of
order a few cm, the condition $P_0>> \Delta P$ (which is necessary in order to be able to
neglect the influence on the contact mechanics from the fluid pressure at the rubber-countersurface)
is satisfied already for loads (at the upper
surface of the cylinder) of order kg.

\vskip 0.3cm

{\bf 10. Summary and conclusion}

Seals are extremely useful devices to prevent fluid leakage. However, 
the exact mechanism of roughness induced leakage is not well understood. 
We have presented a theory of the leak-rate of seals, which is 
based on percolation theory and a recently developed contact mechanics theory.
We have studied both static and
dynamics seals. 
We have presented numerical results for the leak-rate $\dot Q$, and for the lateral size $\lambda_{\rm c}$
and the height $u_{\rm c}$ of the critical constriction. We assumed self affine fractal surfaces and
presented results for how $\dot Q$, $\lambda_{\rm c}$ and $u_{\rm c}$ depend on the root-mean-square
roughness amplitude and the fractal dimension $D_f=3-H$ (where $H$ is the Hurst exponent), and on
the pressure $P_0$ with which the rubber is squeezed against the rough countersurface. 

We have also presented molecular dynamics results 
which show that when two elastic solids with randomly
rough surfaces are squeezed together, as a function of 
increasing magnification or decreasing squeezing pressure, a non-contact channel will percolate 
when relative projected contact area, $A/A_0$, is of order $0.4$, in accordance with 
percolation theory. Finally, we have 
suggested a simple experiment which can be used to test the theory. 

The theory we have presented in this paper is very rough, but we believe that it captures the most important
physics, and that the presented approach can be improved and extended in various ways. 

\vskip 0.3cm

{\bf Acknowledgments:}
We thank Ed Widder (Federal Mogul Sealing Systems) and Matthias Schmidt (IFAS, RWTH Aachen University)
for useful correspondence related to seals. We also thank A. Koenen (Valeo Systeme d'Essuyage) for
very interesting information related to the tribological properties of the contact between rubber and
glass in the context of rubber wiper blades.


\begin{thebibliography}{99}
\bibitem{comment1}
We assume that no cavitation occur at the exit of the critical constriction.
Cavitation may occur if the local pressure at the exit of the critical constriction is negative,
but in this paper we assume so low sliding velocity $v_0$ that this is not the case. If cavitation occur
$p(l_{\rm b}) \approx 0$, or, more accurately, $p(l_{\rm b})$ is 
equal or close to the vapor pressure of the fluid or of gases
dissolved in the fluid. We note that the transition from boundary or mixed lubrication to
hydrodynamic lubrication probably involves cavitation 
at the exit of many narrow constrictions, since
otherwise the total load supported by the fluid film will be very small. It would be very interesting
to study this problem theoretically, since the transition from boundary lubrication to hydrodynamic
lubrication is not well understood.
\bibitem{example}
We did two calculations: For a self affine fractal surface with the rms roughness
of $h_{\rm rms}= 2 \ {\rm \mu m}$ and the squeezing pressure $P_0 = 0.18 \ {\rm MPa}$ we got
$B \approx 0.11$ (and $u_{\rm c} \approx 0.1 \ {\rm \mu m}$ and $\lambda_{\rm c} \approx 16 \ {\rm \mu m}$)
and for $h_{\rm rms}= 4 \ {\rm \mu m}$ and $P_0 = 0.2 \ {\rm MPa}$ we got
$B \approx 0.21$ (and $u_{\rm c} \approx 1 \ {\rm \mu m}$ and $\lambda_{\rm c} \approx 109 \ {\rm \mu m}$).
\bibitem{comment2}
The fluid pressure along the percolation channel increases towards the critical constriction.
Thus, the elastic deformation of the rubber at the critical constriction is determined not just by
the pressure at the constriction but also by the pressure acting on the rubber along the
percolation path. If one assume a strait percolation path and that the pressure along the
percolation path $p(x)=p_{\rm c} x/L$ then one can easily show that 
$\Delta u \approx {\rm ln} (L/\lambda_{\rm c}) \lambda_{\rm c} p_{\rm c} /E$  but the additional
logarithmic factor is never very large
and does not change our qualitative conclusion.
\end{thebibliography}
\end{document}